**Title:**

Contrast-to-noise ratio analysis of microscopic diffusion anisotropy indices in q-space trajectory imaging

**Authors:**

Jan Martin[1]

Sebastian Endt[1]

Andreas Wetscherek[2]

Tristan Anselm Kuder[3]

Arnd Doerfler[4]

Michael Uder[1]

Bernhard Hensel[5]

Frederik Bernd Laun[1]

[1] Institute of Radiology, University Hospital Erlangen, Friedrich-Alexander-Universität Erlangen-Nürnberg (FAU), Erlangen, Germany

[2] Joint Department of Physics, The Institute of Cancer Research and The Royal Marsden NHS Foundation Trust, London, United Kingdom

[3] Department Medical Physics in Radiology, German Cancer Research Center (DKFZ), Heidelberg, Germany

[4] Institute of Neuroradiology, University Hospital Erlangen, Friedrich-Alexander-Universität Erlangen-Nürnberg (FAU), Erlangen, Germany

[5] Center for Medical Physics and Engineering, Friedrich-Alexander-Universität Erlangen-Nürnberg (FAU), Erlangen, Germany

**Correspondence Address:**

Frederik Bernd Laun, Institute of Radiology, University Hospital Erlangen, Maximiliansplatz 3, 91054 Erlangen, Germany. Frederik.Laun@uk-erlangen.de

**Abstract**

Diffusion anisotropy in diffusion tensor imaging (DTI) is commonly quantified with normalized diffusion anisotropy indices (DAIs). Most often, the fractional anisotropy (FA) is used, but several alternative DAIs have been introduced in attempts to maximize the contrast-to-noise ratio (CNR) in diffusion anisotropy maps. Examples include the scaled relative anisotropy (sRA), the gamma variate anisotropy index (GV), the surface anisotropy (UA$_{surf}$), and the lattice index (LI). With the advent of multidimensional diffusion encoding it became possible to determine the presence of microscopic diffusion anisotropy in a voxel, which is theoretically independent of orientation coherence. In accordance with DTI, the microscopic anisotropy is typically quantified by the microscopic fractional anisotropy (µFA).

In this work, in addition to the µFA, the four microscopic diffusion anisotropy indices (µDAIs) µsRA, µGV, µUA$_{surf}$, and µLI are defined in analogy to the respective DAIs by means of the average diffusion tensor and the covariance tensor. Simulations with three representative distributions of microscopic diffusion tensors revealed distinct CNR differences when differentiating between isotropic and microscopically anisotropic diffusion. q-Space trajectory imaging (QTI) was employed to acquire brain in-vivo maps of all indices. For this purpose, a 15 min protocol featuring linear, planar, and spherical tensor encoding was used. The resulting maps were of good quality and exhibited different contrasts, e.g. between gray and white matter. This indicates that it may be beneficial to use more than one µDAI in future investigational studies.

# 1 Introduction

Diffusion-weighted magnetic resonance imaging (DWI) is a central component in stroke and cancer diagnosis [1-4]. Moreover, information on the preferred diffusion directions and the degree of random movement exhibited by water molecules in the tissue can be used to reconstruct a diffusion tensor (diffusion tensor imaging, DTI) and subsequently create maps of diffusional anisotropy [5, 6]. However, diffusion anisotropy indices (DAIs) estimated from data acquired with a traditional single diffusion encoding (SDE) setup [7] suffer from a severe drawback. While the method works well for coherently aligned anisotropic structures, and even allows for advanced methods like fiber tracking [8, 9], differentiating between more complex diffusion characteristics within a voxel remains problematic [10-13]. For example, SDE experiments with isotropic diffusion environments or ensembles of randomly oriented anisotropic diffusion domains result in the same signal attenuation. This limitation may be overcome by encoding two diffusion directions within a single experiment (double diffusion encoding, DDE, [14]) or applying three dimensional diffusion encodings such as magic-angle spinning of the q-vector [15]. Combined with SDE measurements these methods reveal the presence of microscopically anisotropic diffusion compartments, even if they are not well-aligned [14, 16-22]. For example, recent studies found that maps of the microscopic anisotropy correlated well with the characteristic cell geometries of meningioma and glioblastoma, whereas DTI showed little contrast differences [23, 24]. While the technique gains in popularity, efforts are made to find the optimal setup for different organs [25-27], and to keep acquisition protocols within clinically acceptable time limits [28-31].

An important aspect of medical imaging lies in the ease of detecting contrast differences between anatomical structures or between healthy and diseased tissue. Since the advent of DTI in the 1990s [5], a range of DAIs has been introduced in order to achieve the optimal contrast-to-noise ratio (CNR). Fractional anisotropy (FA) and relative anisotropy (RA) were among the first indices introduced to quantify diffusion anisotropy [32].

Papadakis et al. compared FA, RA, and volume ratio (VR) against each other and found FA to feature the highest SNR and greatest detail, while VR had the strongest contrast between areas of low and high anisotropy [33]. Surface and volume anisotropy were introduced by Uluğ et al. as alternative DAIs [34]. They concluded that FA and RA are well suited to distinguish between grey and white matter. Surface anisotropy ($UA_{surf}$) suppressed grey matter and CSF and was preferred to detect differences in a single type of tissue [34]. Armitage et al. presented the gamma variate anisotropy index (GV) as a metric with high contrast and low sensitivity to noise over the anisotropy range commonly found in the human brain [35]. In analytical calculations, Hasan et al. attributed higher SNR to the FA than the RA [36]. An extensive work from Kingsley et al. compared seven DAIs against each other. Here, differences between the metrics were only notable at large anisotropy differences [37].

In most of the recent multidimensional diffusion encoding methods, microscopic diffusion anisotropy is used as an equivalent to the FA, adequately named "microscopic fractional anisotropy" (µFA). In order to calculate the µFA, Westin et al. presented a thorough mathematical framework for generalized diffusion encodings [38]. In our article, we apply this framework to naturally define the microscopic counterparts to several established DAIs, acquire in-vivo maps of these indices, and compare their performance in terms of CNR based on three simulation setups.

## 2 Theory

*2.1 Diffusion anisotropy indices*

Diffusion processes in macroscopic volumes like MRI image voxels may be approximated through a collection of non-exchanging microscopic diffusion environtments [39]. Under the assumption of Gaussian diffusion, each of these micro-domains is described by an individual time-independent diffusion tensor $\boldsymbol{D}$. Hence, the average diffusion tensor corresponding to an entire voxel is defined as $\langle \boldsymbol{D} \rangle$, where $\langle \cdot \rangle$ denotes the ensemble average over all diffusion microenvironments present in the respective voxel. Assuming $\langle \boldsymbol{D} \rangle$ is an axially symmetric tensor, its anisotropy $A$ is defined using the eigenvalues $\lambda_i$ with $\lambda_2 = \lambda_3$ as follows [40]:

$$A = \frac{\lambda_1 - \lambda_2}{\lambda_1 + 2\lambda_2}. \tag{1}$$

Eigenvalues are sorted according to $\lambda_1 \geq \lambda_2$. Additionally, a fourth-order covariance tensor $\mathbb{C}$ may be calculated [38]:

$$\mathbb{C} = \langle \boldsymbol{D}^{\otimes 2} \rangle - \langle \boldsymbol{D} \rangle^{\otimes 2} \tag{2}$$

Here, $\boldsymbol{D}^{\otimes 2} = \boldsymbol{D} \otimes \boldsymbol{D}$ denotes the outer product of $\boldsymbol{D}$ with itself. Westin et al. proposed a framework to estimate diffusion anisotropy indices based on $\langle \boldsymbol{D} \rangle$ and $\mathbb{C}$ in combination with the isotropic fourth-order tensors $\mathbb{E}_{\text{shear}}$, $\mathbb{E}_{\text{bulk}}$, and $\mathbb{E}_{\text{iso}}$ [38, 41].

$$\mathbb{E}_{\text{shear}} = \frac{1}{9}\begin{pmatrix} 2 & -1 & -1 & 0 & 0 & 0 \\ -1 & 2 & -1 & 0 & 0 & 0 \\ -1 & -1 & 2 & 0 & 0 & 0 \\ 0 & 0 & 0 & 3 & 0 & 0 \\ 0 & 0 & 0 & 0 & 3 & 0 \\ 0 & 0 & 0 & 0 & 0 & 3 \end{pmatrix} \qquad (3)$$

$$\mathbb{E}_{\text{bulk}} = \frac{1}{9}\begin{pmatrix} 1 & 1 & 1 & 0 & 0 & 0 \\ 1 & 1 & 1 & 0 & 0 & 0 \\ 1 & 1 & 1 & 0 & 0 & 0 \\ 0 & 0 & 0 & 0 & 0 & 0 \\ 0 & 0 & 0 & 0 & 0 & 0 \\ 0 & 0 & 0 & 0 & 0 & 0 \end{pmatrix} \qquad (4)$$

$$\mathbb{E}_{\text{iso}} = \mathbb{E}_{\text{shear}} + \mathbb{E}_{\text{bulk}} \qquad (5)$$

In this context the FA, the same DAI commonly used in DTI [32], is calculated according to the following equation:

$$FA^2 = \frac{3}{2}\frac{<\langle \boldsymbol{D}\rangle^{\otimes 2},\ \mathbb{E}_{\text{shear}}>}{<\langle \boldsymbol{D}\rangle^{\otimes 2},\ \mathbb{E}_{\text{iso}}>}. \qquad (6)$$

Here, the brackets $<\mathbb{A}, \mathbb{B}>$ denote the inner product of two fourth-order tensors $\mathbb{A}$ and $\mathbb{B}$. An important limitation of the FA is the entanglement of anisotropy and orientation coherence of the microscopic diffusion tensors, as implied by the outer product of the already averaged diffusion tensor $\langle \boldsymbol{D}\rangle$ with itself in eq. 6 [12, 38].

However, when averaging over the outer product of the microscopic diffusion tensors instead, one can compute a measure for the microscopic fractional anisotropy [38]:

$$\mu FA^2 = \frac{3}{2}\frac{<\langle \boldsymbol{D}^{\otimes 2}\rangle, \mathbb{E}_{\text{shear}}>}{<\langle \boldsymbol{D}^{\otimes 2}\rangle, \mathbb{E}_{\text{iso}}>}. \tag{7}$$

Compared to the FA, this metric is independent of the alignment between the micro-domains, and effectively disentangles the effects of anisotropy and orientation coherence [16].

Several additional DAIs have been proposed to quantify the anisotropy of the diffusion tensor [40]. The relative anisotropy was introduced in the same work as the FA [32]. Normalizing the range of RA to one results in the scaled relative anisotropy (sRA). Similar to the FA in eq. 6, this DAI may also be expressed in terms of fourth-order tensors.

$$sRA^2 = \frac{1}{2}\frac{<\langle \boldsymbol{D}\rangle^{\otimes 2}, \mathbb{E}_{\text{shear}}>}{<\langle \boldsymbol{D}\rangle^{\otimes 2}, \mathbb{E}_{\text{bulk}}>} \tag{8}$$

For axially symmetric $\langle \boldsymbol{D}\rangle$ the relationship $sRA = A$ holds (see eq. 1). In conformity with eq. 7, the microscopic sRA (μsRA) is defined as follows:

$$\mu sRA^2 = \frac{1}{2}\frac{<\langle \boldsymbol{D}^{\otimes 2}\rangle, \mathbb{E}_{\text{shear}}>}{<\langle \boldsymbol{D}^{\otimes 2}\rangle, \mathbb{E}_{\text{bulk}}>}. \tag{9}$$

Adapted from the sRA, Armitage and Bastin derived the gamma variate anisotropy index to improve the sensitivity over the common sRA range found in the human brain [35]. GV can be computed directly from the sRA.

$$GV = 259.57[1 - \exp(-8 \cdot sRA)(32 \cdot sRA^2 + 8 \cdot sRA + 1)]/256 \tag{10}$$

The microscopic equivalent, μGV, is defined with the same formula, using the μsRA instead.

$$\mu GV = 259.57[1 - \exp(-8 \cdot \mu sRA)(32 \cdot \mu sRA^2 + 8 \cdot \mu sRA + 1)]/256 \tag{11}$$

Ulŭg et al. proposed a surface anisotropy index based on additional rotational invariants of the diffusion tensor [34]. In tensor notation, $UA_{surf}$ may be calculated as follows.

$$UA_{surf} = 1 - \sqrt{1 - \frac{<\langle \boldsymbol{D} \rangle^{\otimes 2}, \mathbb{E}_{shear}>}{2 <\langle \boldsymbol{D} \rangle^{\otimes 2}, \mathbb{E}_{bulk}>}} \qquad (12)$$

Again, the microscopic surface anisotropy can be obtained by averaging over the outer product of the microscopic tensors instead.

$$\mu UA_{surf} = 1 - \sqrt{1 - \frac{<\langle \boldsymbol{D}^{\otimes 2} \rangle, \mathbb{E}_{shear}>}{2 <\langle \boldsymbol{D}^{\otimes 2} \rangle, \mathbb{E}_{bulk}>}} \qquad (13)$$

The lattice index (LI) was originally introduced as an intervoxel measure for diffusion anisotropy [42]. To this end, the LI was calculated as the weighted average of the eight surrounding voxels in a slice. Kingsley et al. proposed an alternative expression of LI for a single voxel expressed in terms of the FA [40]:

$$LI = (FA + FA^2)/2 \qquad (14)$$

Similar to eq. 11, the microscopic lattice index (µLI) is calculated directly from the µFA.

$$\mu LI = (\mu FA + \mu FA^2)/2 \qquad (15)$$

All DAIs and µDAIs presented here range from zero in case of isotropic diffusion to one for completely anisotropic microscopic diffusion tensors. Besides the DAIs discussed in this section, anisotropy indices exist that are calculated from the third power of eigenvalues of $\langle \boldsymbol{D} \rangle$ such as the volume ratio [32] and volume anisotropy [34].

However, the QTI formalism is ill-suited to translate such indices to microscopic indices, because solely the variance $\langle \mathbf{D}^{\otimes 2} \rangle$ is estimated. By definition $\langle \mathbf{D}^{\otimes 2} \rangle$ depends on eigenvalues to the second power, but offers no information on the third power. Hence, DAIs calculated from the third power of eigenvalues were not considered in this article.

*2.2 q-Space trajectory imaging*

Assuming Gaussian diffusivity in all diffusion micro-domains, the measured signal $S$ only depends on the applied b-tensor $\mathbf{B}$ [38].

$$S(\mathbf{B})/S_0 = \langle exp(-<\mathbf{B},\mathbf{D}>) \rangle \tag{16}$$

Here, $S_0$ corresponds to the signal intensity without diffusion weighting. Similar to diffusion kurtosis imaging (DKI, [43]), fitting $\langle \mathbf{D} \rangle$ and $\mathbb{C}$ requires a second-order expansion of the signal equation [38].

$$S(\mathbf{B})/S_0 \approx exp\left(-<\mathbf{B},\langle \mathbf{D}\rangle> + \frac{1}{2}<\mathbf{B}^{\otimes 2},\mathbb{C}>\right) \tag{17}$$

Note, that the inversion of eq. 17 requires measurements with b-tensors of rank 2 or higher. Consequently, it is not possible to estimate diffusional variance from SDE [44] experiments alone [38].

# 3 Methods

*3.1 Image acquisition*

In this work, linear, planar, and spherical b-tensors were combined to achieve the multidimensional diffusion encoding necessary for QTI. Time-varying diffusion gradients were applied to generate higher order b-tensors. To minimize the required echo time, the gradient trajectories were calculated numerically in MATLAB (The MathWorks, Natick, MA, USA). The optimization algorithm was set up according to the method proposed by Sjölund et al. [45] and accounted for the following constraints: Maximum diffusion gradient amplitude limited to 80 mT/m, restricted by the available gradient hardware. The allowed slew rate was capped at 100 mT/ms to avoid peripheral nerve stimulation. During the 180° refocusing pulse, all diffusion gradients were turned off. Linear and planar b-tensors were optimized under the Euclidean L2 norm, whereas a maximum norm was employed for the spherical b-tensor.

Image acquisition was carried out with a single-shot spin echo EPI sequence. All data was measured on a 7T system (Magnetom Terra, Siemens Healthcare GmbH, Erlangen, Germany) using a 1Tx/32Rx head coil (Nova Medical, Wilmington, USA). The scanner was equipped with 80 mT/m gradients with a maximum slew rate of 200 mT/ms. Imaging parameters were set up as follows: TE 80 ms, TR 3500 ms, field of view 230 x 230 mm², voxel size 2x2x4 mm³, 20 slices, acquisition bandwidth 1512 Hz/pixel, phase partial Fourier factor 0.75, in-plane acceleration factor 3 using GRAPPA reconstruction.

The protocol included linear, planar, and spherical b-tensors at five b-values (0, 100, 500, 1000, 1500, 2000 s/mm²). Linear and planar b-tensors were rotated in 16 noncollinear directions as described in [46]. Spherical b-tensors were averaged five times over three orthogonal directions to minimize potential inconsistencies resulting from diffusion time anisotropy [47].

Total acquisition time for the diffusion weighted images was 15:46 min, with 235 individual measurements per slice.

The local institutional review board approved the study protocol. A healthy volunteer (age 21) was recruited, whose written informed consent was obtained prior to scanning.

*3.2 Data analysis*

Images were corrected for motion and eddy-current artifacts with MATLAB and ElastiX [48]. The applied routine matched uncorrected images with reference images extrapolated from data acquired at b ≤ 1000 s/mm² [49]. In order to reduce Gibb's ringing artifacts, smoothing was carried out with a Gaussian filter with a standard deviation of 0.5 voxel. The average diffusion tensor $\langle D \rangle$ and the covariance tensor $\mathbb{C}$ were fitted voxelwise according to eq. 17 [38]. DAIs and µDAIs were calculated following the framework presented in the theory section of this article. Values above one or below zero were set to one and zero, respectively.

Direction encoded maps of the FA and µFA maps were used to define volumetric regions-of-interest (RoIs) within the corpus callosum (CC), the posterior limb of the capsula interna (CI), the thalamus (TH), as well as the frontal ventricle (VE). RoIs were placed bilaterally if possible (see Fig. 1, [50]).

*3.3 Simulations*

In order to investigate the contrast-to-noise ratio (CNR) of the µDAIs, numerical simulations with three artificial diffusion tensor distributions (DTDs) were carried out. The three DTDs consisted of an anisotropic compartment with coherent orientation combined with one of the following compartments (see Fig. 2):

a) an anisotropic compartment of coherent orientation rotated by an angle $\alpha$ compared to the first compartment (crossing angle)

b) an anisotropic compartment with random orientations

c) an isotropic compartment

The relative weight between the first and the second compartment was $f$ and $1-f$, respectively, with $0 \leq f \leq 1$. In setup a), $f$ was fixed at 50% (two microscopic diffusion environments) and $\alpha$ was varied between 0° and 90°. The second compartment in setup b) consisted of $10^4$ diffusion micro-domains with random orientations. Setup c) contained 99 individual diffusion tensors. For setup b) and c), $f$ was varied between 0 and 100%. Setup a) was intended to replicate a voxel containing crossing fibers. The second and third setup are roughly similar to the fiber structure found in grey matter [5, 51], white matter mixed with cerebrospinal fluid (CSF), or cancerous infiltrations in white matter [23, 24].

The computations were carried out in MATLAB. Diffusion properties were derived from the in-vivo data: To obtain average diffusion values for white matter, a region-of-interest (RoI) was defined by thresholding the µFA at 0.75 (see Fig. 1). Based on maps of the mean diffusivity (MD) and µFA, the axial diffusivity was chosen to be AD = 2.04 µm²/ms, and the radial diffusivity set to RD = 0.26 µm²/ms for all anisotropic diffusion tensors in the simulations. In setup c), the MD of the isotropic share was set to 0.85 µm²/ms, equal to the MD of the anisotropic compartment. The resulting µDAI values at $\alpha = 0°$ and $f = 0$ (µDAI$_0$) were µFA = 0.86, µsRA = 0.70, µGV = 0.93, µUA$_{surf}$ = 0.29, and µLI = 0.80.

A synthetic diffusion weighted signal was calculated according to eq. 16 [38]. b-Tensor shapes and b-values replicated the in-vivo protocol described above. The initial signal intensity $S_0$ and the standard deviation for the Gaussian noise $\sigma_{noise}$ was estimated based on two unweighted images following the difference method described by Dietrich et al. [52]. Subsequently, $S_0$ was set to 154.9, and $\sigma_{noise}$ to 2.5, resulting in an effective SNR of 61.96.

Average diffusion and covariance tensors were fitted to the artificial signal curves following eq. 17. The process was repeated $10^6$ times for each setup. CNRs were calculated with respect to the tensor distribution at $\alpha = 0$ and $f = 0$ respectively, indicated here by the subscript '0':

$$CNR = \frac{|\langle \mu DAI \rangle - \langle \mu DAI_0 \rangle|}{\sqrt{(\sigma^2 + \sigma_0^2)/2}}. \tag{18}$$

In this case, $\langle \mu DAI \rangle$ represents the average of the respective µDAI over all repetitions, and $\sigma$ denotes the corresponding standard deviation of the µDAI.

In order to investigate the impact of limited SNR on the estimation of µDAIs, an additional set of simulations was carried out. The three diffusion tensor distributions from setups a), b), and c) (see Fig. 2) were fixed at $\alpha = 90°$ and $f = 50\%$, respectively. The value of $\sigma_{\text{noise}}$ was chosen to replicate SNRs of 10, 15, 20, 25, 30, 35, 40, 45, and 50. As before, each simulation was repeated $10^6$ times.

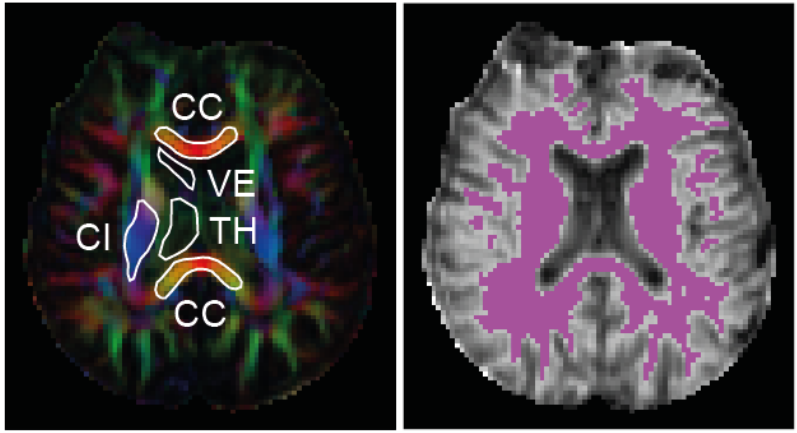

**Figure 1:** Representative maps of the direction encoded FA (left) and the μFA (right). The RoIs evaluated in this article are drawn in white. CI, VE, and TH are only shown in the left half of the brain to benefit visibility, but were evaluated bilaterally. A violet region indicates the RoI defined by thresholding μFA at 0.75 to evaluate the average diffusion properties of white matter and estimate the SNR used in the numerical simulations.

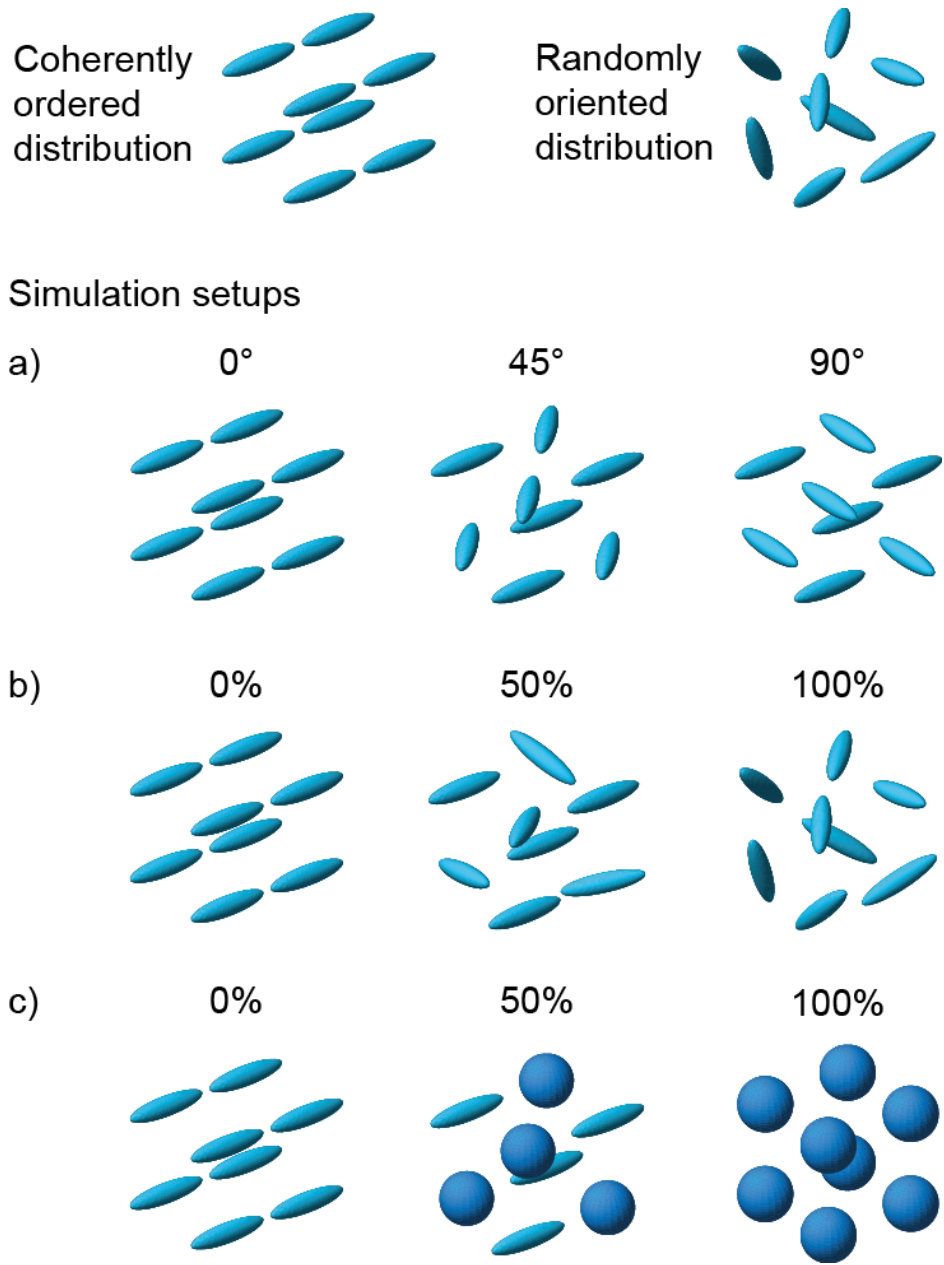

**Figure 2:** Overview of diffusion tensor distributions. Three setups were used for the numerical simulations: a crossing fiber setup (a), a combination of well-aligned and randomly oriented compartments (b), and a mixture of anisotropic and isotropic diffusion (c). The number of depicted tensors is limited to eight in order to improve visibility.

## 4 Results

Figure 3 shows the dependence of five DAIs and their microscopic equivalents on the anisotropy A (see eq. 1). The indices were computed with $\langle D \rangle$ and $\mathbb{C}$ (see eq. 2) of the respective DTD using eqs. 6 to 15. In the anisotropic coherently ordered distribution (see Fig. 2), the microscopic diffusion domains are well-aligned; DAI and µDAI follow the same trend and replicate the findings by Kingsley et al. [37]. In a randomly oriented anisotropic distribution, the anisotropy of the average diffusion tensor $\langle D \rangle$ equals zero. As a result, every DAI is close to zero for such a DTD, while the µDAIs show no discernable difference to the well-aligned case.

Figure 4 displays the normalized µDAIs in a setting without noise, as well as the normalized µDAIs, standard deviation $\sigma$, and CNR at an SNR level equivalent to the in-vivo acquisitions.

Simulation results for the setups a) and b) show a distinct dependency of the normalized µDAI on the crossing angle $\alpha$ and the percentage of randomly oriented anisotropic domains $f$, respectively. The effect is intensified with noisy data, and is most pronounced for µUA$_{surf}$. Resulting CNR curves, however, are almost identical for all considered µDAIs.

In setup c), where parts of the anisotropic distribution were replaced by isotropic diffusion environments, all microscopic anisotropy metrics show a monotonic decrease. At finite SNR, the µDAIs do not decrease to zero even if the entire voxel contains only isotropic domains. This effect is caused by limiting the minimum value of the µDAIs to zero. Compared to setup a) and b), the overall CNR is higher in c). At high average anisotropy, all µDAIs achieve similar CNRs. For distributions with approximately equal parts of anisotropic and isotropic microenvironments, µLI and µsRA have a slightly higher CNR than the other indices. If the percentage of isotropic micro-domains exceeds 90%, µGV shows a rapid increase in CNR, surpassing the rest of the µDAIs.

Simulation results for different levels of SNR are shown in Fig. 5. In all three investigated setups (see Fig. 2), the estimated mean µDAI values showed negligible SNR dependence down to a SNR of roughly 20 to 30. Towards even lower SNR, the µDAIs were consistently underestimated. For setup a) and b), the underestimation was most pronounced for µUA$_{surf}$, and lest severe for the µFA. For setup c), µUA$_{surf}$ showed the smallest underestimation and µGV the largest. The coefficient of variation (CV) increased towards lower SNR. In setups a) and b), CV was highest for µUA$_{surf}$ and lowest for µFA and µGV. In setup c) the CV was overall larger than in setups a) and b). Here, µUA$_{surf}$ had the largest CV and µFA had the lowest CV. In general, µFA, µsRA, and µLI show comparable CNR(SNR) dependencies. The CNR is lowest for µUA$_{surf}$ in setup a) and c), and lowest for µGV in setup b) for SNR > 20. Above SNR = 20, the CNR(SNR) curve is primarily affected by CV, which depends roughly inversely on SNR, so that CNR(SNR) increases approximately linearly (see eq. 18). Below SNR = 20, both, µDAI as well as CV, affect the CNR, and a derivation from the linear behavior is observed.

In-vivo maps for all DAIs and µDAIs discussed in this article are shown in Fig. 6. Anisotropy values for the RoIs in the CC, CI, TH, and VE are listed in Table 1.

In general, highly coherent structures such as CC and CI (see Fig. 1) appear bright in every map. When comparing DAIs to µDAIs, the dependence on orientation coherence becomes apparent. Sections of incoherently ordered white matter (see arrows in Fig. 6) or gray matter are heavily attenuated in all DAI maps, while they retain similar intensities to the coherently ordered structures in most of the µDAI maps. For the chosen windowing parameters, µFA appears homogeneous in white matter with a continuous decrease of brightness when going from central white matter regions to outer regions. The intensity difference between outer white matter and gray matter appears to be relatively smooth.

In comparison, µsRA displays a sharper drop in brightness when going from central to outer white matter regions, indicating a wider dynamic range. µGV appears rather homogeneous over the entire white matter, grey matter appears dark. µUA$_{surf}$ shows CC and CI very bright, while outer white matter structures and grey matter are attenuated heavily. Finally, µLI appears very similar to µFA, with a seemingly wider dynamic range.

Numerical values of all µDAIs in white matter are larger than those of the corresponding DAI (see Table 1), and to an even larger extent in gray matter. In the RoI in the ventricles, DAIs and µDAIs are both close to zero and almost identical.

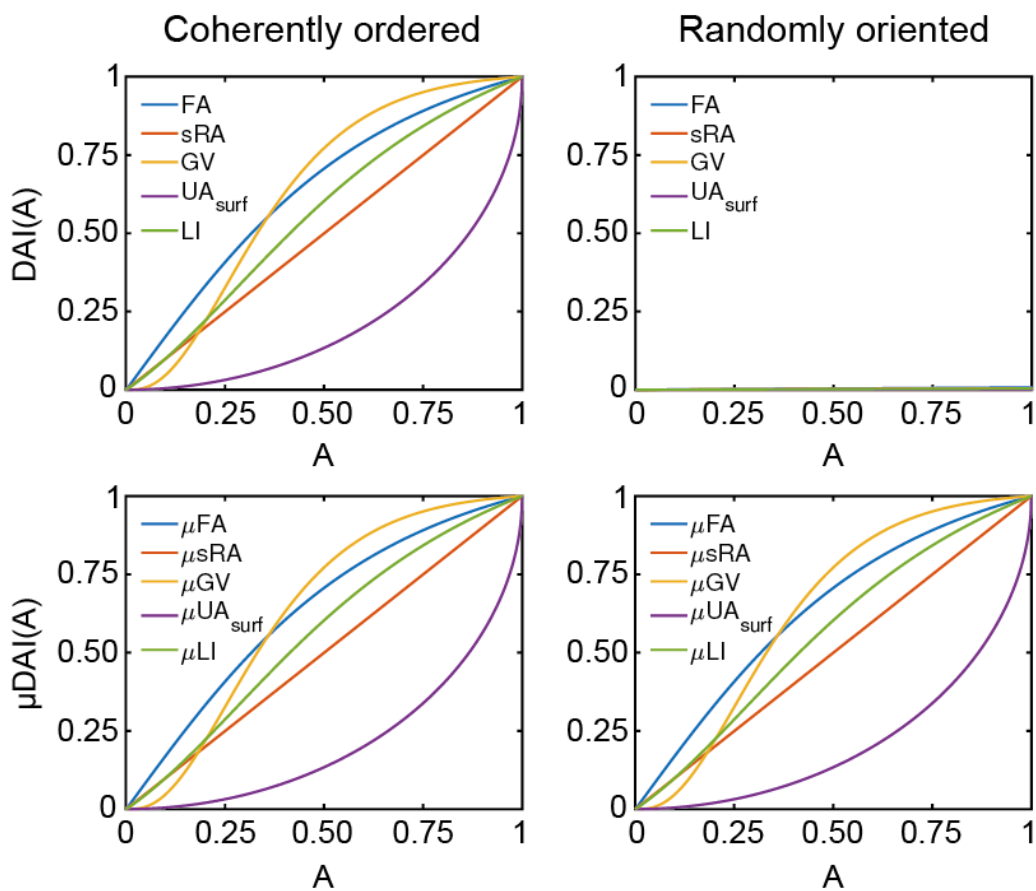

**Figure 3:** Dependence of the DAIs and µDAIs discussed in this article on the anisotropy A, which ranges from zero (isotropic diffusion) to one (linear diffusion). The DAIs are affected heavily by the reduced orientation coherence, whereas all µDAIs remain undisturbed.

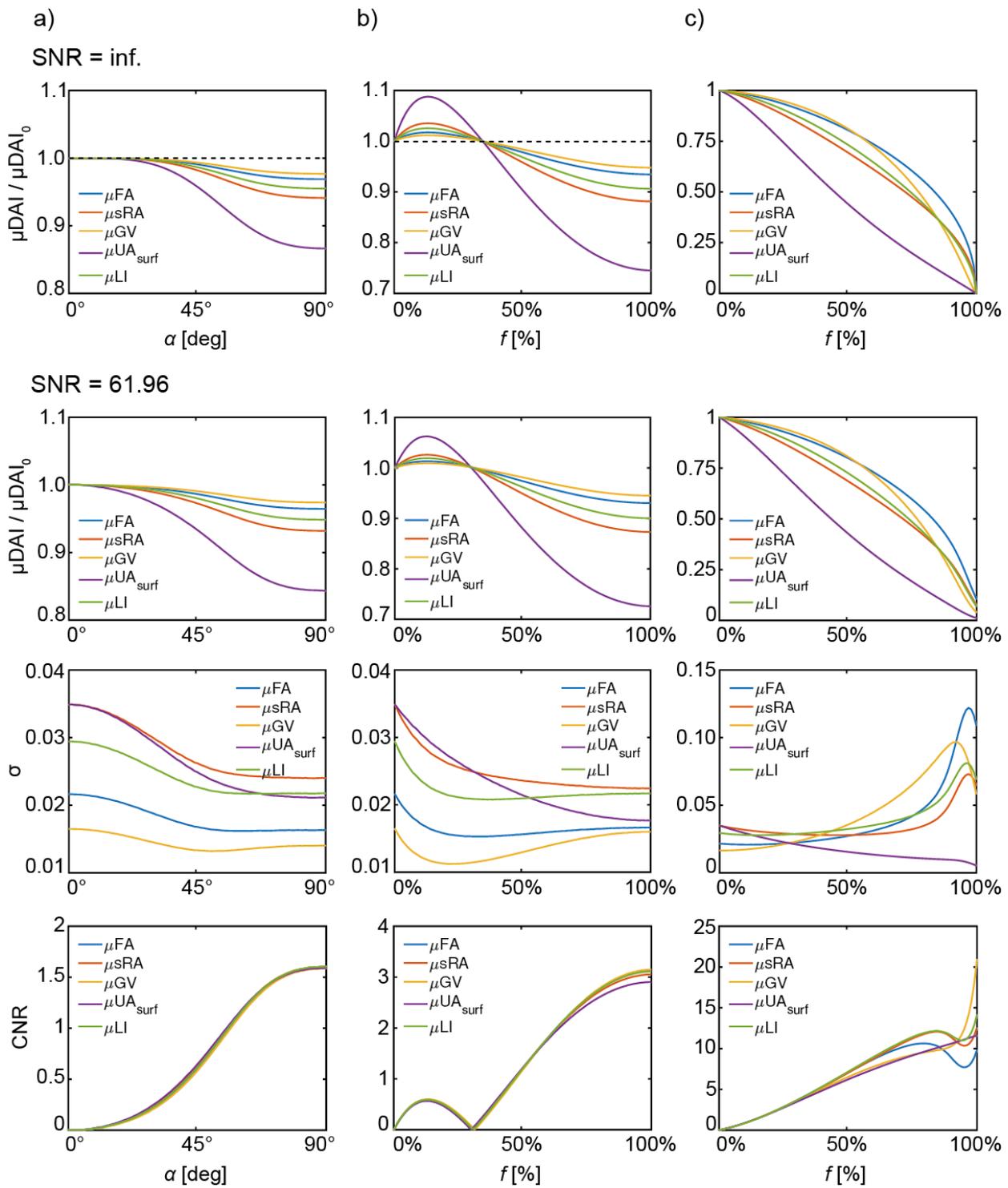

**Figure 4:** CNR simulation results for the setups shown in Fig. 2. Upper two rows: normalized µDAIs at infinite and finite SNR. Third row: standard deviation $\sigma$ over $10^6$ repetitions. Bottom row: CNR (see eq. 18). A dashed black line indicates the normed µDAIs if calculated directly from $\langle \boldsymbol{D} \rangle$ and $\mathbb{C}$ of the tensor distribution.

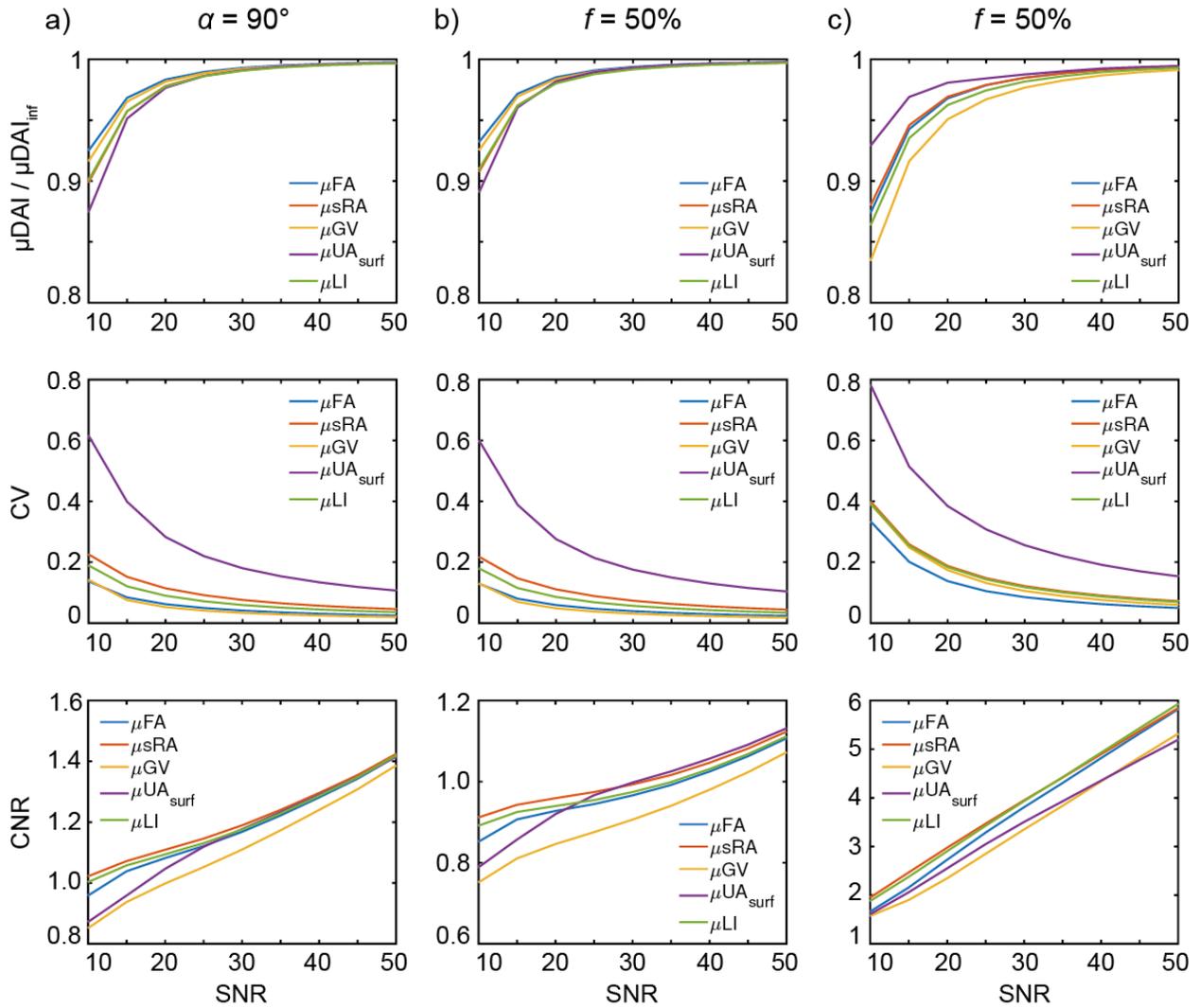

**Figure 5:** SNR dependency of different µDAIs. Displayed are the average µDAIs normalized to their respective values at infinite SNR µDAI$_{inf}$ (top row), the coefficient of variation CV (middle row), and the CNR at different levels of SNR (bottom row). Setup a) was fixed at $\alpha = 90°$. Setup b) and c) at $f = 50\%$ (see Fig. 2). For each setup the simulations were repeated $10^6$ times.

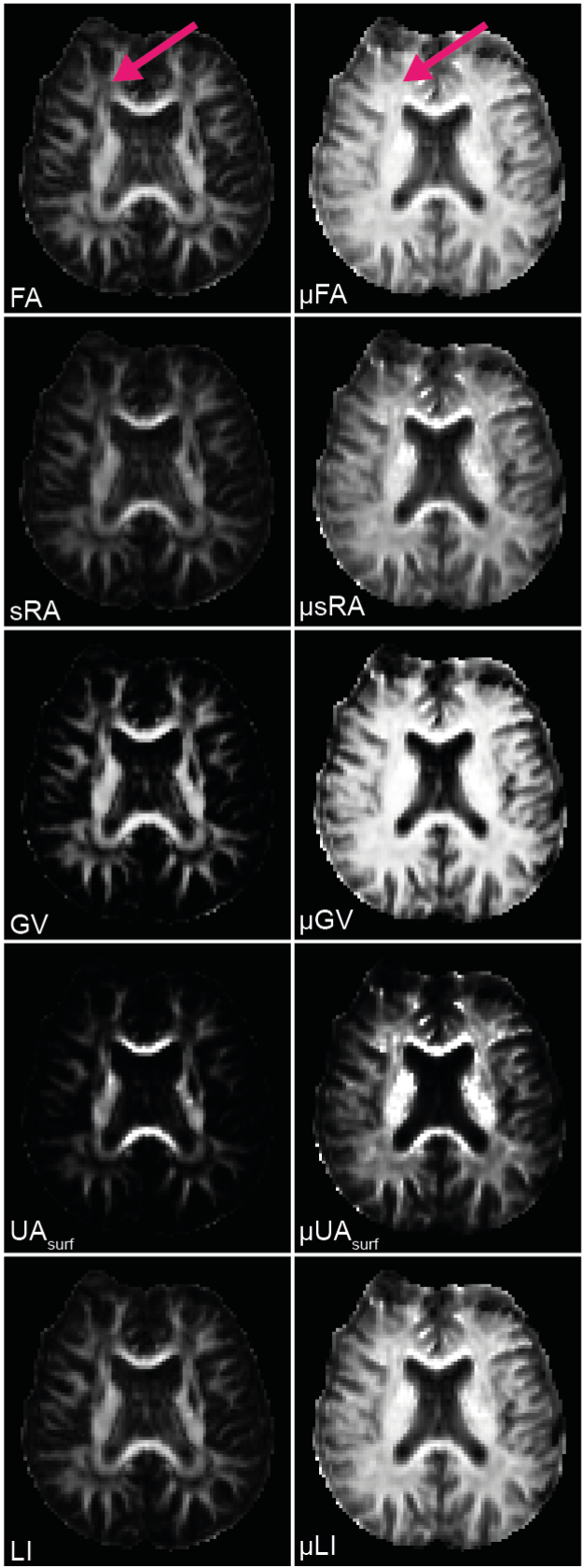

**Figure 6:** In-vivo maps of the DAIs and µDAIs discussed in this article. A single slice in transversal orientation is shown. Arrows indicate a region of crossing white matter fibers. All contrasts were calculated from the same set of diffusion-weighted data, acquired with linear, planar, and spherical b-tensor encoding. Windowing: 0 to 1, except $UA_{surf}$ 0 to 0.25, and $\mu UA_{surf}$ 0 to 0.5.

**Table 1:** Anisotropy values corresponding to the DAIs and µDAIs in Fig. 6 for the RoIs shown in Fig. 1. Values are presented as mean ± standard deviation over the voxels contained in the respective RoI.

|  | CC | CI | TH | VE |
|---|---|---|---|---|
| FA | 0.73 ± 0.14 | 0.67 ± 0.11 | 0.320 ± 0.083 | 0.131 ± 0.052 |
| sRA | 0.55 ± 0.16 | 0.48 ± 0.13 | 0.193 ± 0.054 | 0.076 ± 0.031 |
| GV | 0.78 ± 0.17 | 0.71 ± 0.14 | 0.21 ± 0.11 | 0.031 ± 0.031 |
| $UA_{surf}$ | 0.20 ± 0.15 | 0.14 ± 0.14 | 0.020 ± 0.011 | 0.0034 ± 0.0028 |
| LI | 0.65 ± 0.17 | 0.57 ± 0.14 | 0.214 ± 0.069 | 0.075 ± 0.034 |
| µFA | 0.89 ± 0.11 | 0.946 ± 0.042 | 0.803 ± 0.068 | 0.131 ± 0.049 |
| µsRA | 0.78 ± 0.18 | 0.868 ± 0.090 | 0.624 ± 0.097 | 0.076 ± 0.023 |
| µGV | 0.846 ± 0.093 | 0.921 ± 0.022 | 0.726 ± 0.065 | 0.075 ± 0.029 |
| $µUA_{surf}$ | 0.49 ± 0.31 | 0.56 ± 0.22 | 0.230 ± 0.095 | 0.0033 ± 0.0026 |
| µLI | 0.93 ± 0.14 | 0.978 ± 0.059 | 0.872 ± 0.089 | 0.030 ± 0.031 |

## 5 Discussion

When comparing DAIs in terms of CNR, Kingsley et al. reported that DAIs estimated from the second power of eigenvalues of the diffusion tensor, e.g. FA, sRA, GV, and UA$_{surf}$, performed almost identical over the entire anisotropy range [37]. For the complementary µDAIs, in the clinically most relevant setup c), we observed that µsRA and µLI obtained the highest CNR over a wide range of anisotropy up to $f \approx 90\%$, where µGV obtained the best CNR (see Fig. 4). Furthermore, Kingsley et al. also stated that DAIs based on the third power of eigenvalues achieved a lower CNR overall. These DAIs were not considered as possible candidates for µDAIs in this work, as generalizing them in the scope of the QTI framework is difficult.

Considering that diffusion microstructure imaging is currently primarily applied to investigate white matter in the brain [16, 21, 23, 24, 26, 29, 38, 45, 47, 53, 54], the simulation results with setup c) suggest that either µLI or µsRA provide an improvement in CNR over the other µDAIs (see Fig. 4), while µUA$_{surf}$ and µGV seem less recommendable at lower SNR (see Fig. 5). Remarkably, µLI is derived directly from the µFA (see eq. 15), but outperforms it in terms of CNR at medium anisotropy values (see Fig. 4). This indicates that it is possible to optimize the CNR of anisotropy maps through basic mathematical operations.

Several works investigated the impact of low SNR on DAIs. An early study from Pierpaoli and Basser revealed a dependency of the eigenvalues of $\langle D \rangle$ on the SNR, as FA increased towards lower SNR [42]. Similarly, Bastin et al. reported an increase in RA at SNRs below 50 [55]. Simulations from Jones and Basser showed that the FA changes with decreasing SNR but whether the theoretical value is over- or underestimated depends on the maximum b-values used during data acquisition [56]. In our simulations, µDAIs were increasingly underestimated at SNRs below 30 (see Fig. 5). Through all investigated simulation setups in this work, µFA showed the smallest CV at low SNR, with less underestimation than most of the other µDAIs.

At SNR levels below 30, the CNR in setup c) at $f = 50\%$ was highest for µsRA and µLI (see Fig. 5). Thus, if the measurement protocol does not allow for good SNR values, the subsequent loss in CNR can be reduced by choosing an appropriate µDAI.

The simulation results also show that the SNR of the in-vivo data (SNR = 61.69) was sufficiently high to avoid major bias.

Microscopic diffusion anisotropy indices are intended to separate the influence of anisotropy and orientation coherence on macroscopic metrics such as the FA [16, 21]. For the simulation setups a) and b), theory predicts no dependency of any µDAI on neither $\alpha$ nor $f$. The anisotropy of the microscopic diffusion tensors remained the same during the simulations, and an ideal µDAI would be expected to yield a constant value for all settings. This holds true when calculating the µDAI with $\langle \boldsymbol{D} \rangle$ and $\mathbb{C}$ known from the tensor distribution (see Fig. 4, dashed line). However, numerical results achieved with QTI indicate that, to a varying degree, each µDAI considered in this article exhibited a dependency on $\alpha$ and $f$. This discrepancy between theory and simulation arises from the fact that QTI approximates the diffusion-weighted signal with a second-order cumulant expansion (see eq. 17). Yet, higher order terms have a non-negligible influence, and cause the observed dependency of the µDAIs on orientation coherence. In the simulation performed here, this error remained below 10% for most µDAIs while reaching a maximum value of more than 20% for µUA$_{surf}$. Following this line of thought, small changes in in-vivo µDAI maps not necessarily reflect a difference in microscopic anisotropy, but could also be attributed to a different alignment. Alternative reconstruction methods which take higher order terms into account [16, 53, 54] may improve the accuracy but could exhibit a different response to noise than QTI. A thorough comparison between different approaches is advisable but was beyond the scope of this work.

Compared to the other two simulation setups, the CNR in setup c) is remarkably higher for all µDAIs (see Fig. 4). This indicates that a major decrease in microscopic anisotropy would still be detectable with reasonable accuracy and precision even in the presence of a possible bias introduced by the second-order approximation of the QTI framework.

Several limitations have to be taken into account. Actual in-vivo diffusion tensor distributions may be far more complicated than the simplified compositions used for our simulations. Yet, this work was not intended to provide an extensive biological model, but instead to suggest the use of alternative indices for microscopic anisotropy, and to depict the possible difference in CNR for simple cases such as crossings or decreasing average anisotropy. A further limitation is that the simulation parameters were based on brain white matter. As microscopic diffusion imaging gains in popularity, other regions such as prostate or kidneys become potential fields of research [25, 27]. The associated differences in diffusivity and acquisition parameters may lead to changes in CNR. Finally, although CNR provides an objective measure for image contrast, the choice for a µDAI in clinical routine or research will ultimately depend on the subjective performance of the physician or scientist when interpreting the images. The observed differences in visual appearance of the maps, as well as the results of the CNR simulations, suggest that investing time to find the best-suited µDAI for a given research question is helpful to improve subsequent data evaluation.

## 6 Conclusion

We introduced several alternatives to the µFA by translating known macroscopic DAIs. Simulations revealed a dependency of all µDAIs on orientational coherence, introduced by a second-order approximation of the signal equation. However, this effect was small compared to the µDAI changes over the entire anisotropy range. While all µDAIs discussed in this article seem well suited for QTI analysis, µsRA and µLI generally provided the highest CNR when differentiating between isotropic and microscopically anisotropic diffusion. In-vivo maps were stable but visually different, indicating that the proper choice of µDAI may be beneficial in future clinical studies.

## Acknowledgements

Financial support by the Deutsche Forschungsgemeinschaft is gratefully acknowledged (grant number DFG LA 2804/6-1).

**Appendix**

In order to explain the remarkable similarities in CNR of the different µDAIs in Fig. 4, it is useful to express µFA (see eq. 7) and µUA$_{surf}$ (see eq. 13) in terms of the µsRA:

$$\mu FA = \mu sRA \cdot \sqrt{3/(2 \cdot \mu sRA^2 + 1)} \qquad (A.1)$$

$$\mu UA_{surf} = 1 - \sqrt{1 - \mu sRA^2} \qquad (A.2)$$

Figure 7 shows the dependency of µFA, µGV, µUA$_{surf}$, and µLI on the µsRA. In simulation setups a) and b), as well as in setup c) at $f \leq 30\%$, the µsRA ranges between 0.7 and 0.6. A Taylor series expansion at µsRA = 0.7 up to first order shows that the remaining µDAIs are well approximated by a linear function $f(\mu sRA) = c_1 \cdot \mu sRA + c_2$ in this range (see Fig. 7, dashed line). Substituting the mean $\langle f(\mu sRA)\rangle = c_1 \langle \mu sRA\rangle + c_2$ and standard deviation $\sigma(f(\mu sRA)) = |c_1| \cdot \sigma(\mu sRA)$ in eq. 18, it becomes clear that all µDAIs achieve similar CNR as long as the linear approximation is valid. Only at sufficiently big differences between the microscopic anisotropy and the reference point for the CNR calculation do the CNR curves of the various µDAIs begin to differ from each other (see Fig. 4, setup c)). It is noteworthy that the CNR curves depend on the reference point µDAI$_0$. In this article we chose an undisturbed anisotropic distribution of micro-domains as the reference point, as it seemed most appropriate for brain imaging.

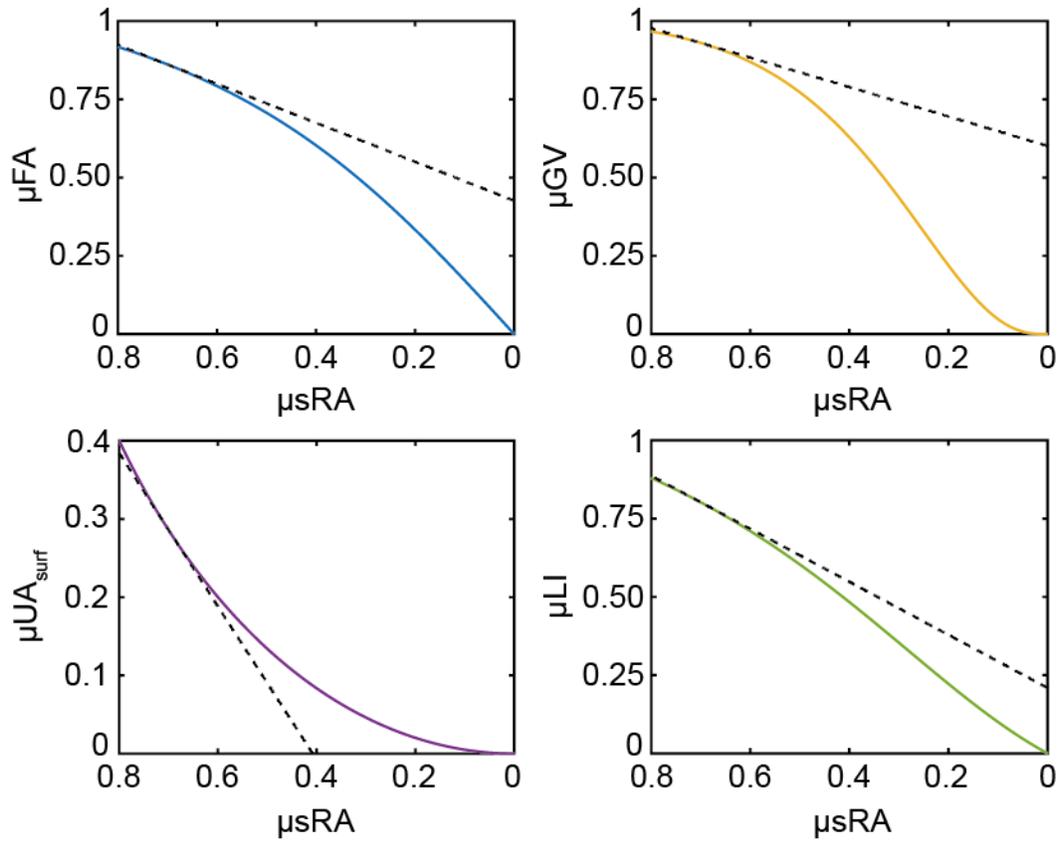

**Figure 7:** µDAIs discussed in this article in terms of the µsRA (eqs. 11, 14, A.1, A.2). A dashed line indicates the first-order Taylor series expansion of the respective µDAI around µsRA = 0.7.